\documentclass[aps,prb,reprint,superscriptaddress, longbibliography]{revtex4-2}
\usepackage{graphicx,color}
\usepackage{amssymb}   
\usepackage{amsmath}

\usepackage{epstopdf}
\usepackage{natbib}
\usepackage{tabularx,booktabs}
\usepackage[colorlinks=true, letterpaper=true, pdfstartview=FitV, linkcolor=red, citecolor=blue, urlcolor=blue]{hyperref}
\usepackage{bm}
\usepackage{color}
\usepackage{braket}
\usepackage{hyperref}
\usepackage{url}
\usepackage{xcolor}

\begin{document}
\title{Floquet Topological Spin-Valley-Layertronics on a Layered Dice Lattice}

\author{Jianqi Zhong}
\affiliation{Thrust of Advanced Materials \& Quantum Science and Technology Center, The Hong Kong University of Science and Technology (Guangzhou), 1 Duxue Rd., Nansha, Guangzhou, China}

\author{Teng-Fei Ying}
\affiliation{Thrust of Advanced Materials \& Quantum Science and Technology Center, The Hong Kong University of Science and Technology (Guangzhou), 1 Duxue Rd., Nansha, Guangzhou, China}

\author{Jinyu Zou}
\email{jyzou@hust.edu.cn}
\affiliation{Wuhan National High Magnetic Field Center \& School of Physics, Huazhong University of Science and Technology, Wuhan, China}

\author{Benjamin T. Zhou}
\email{tongz@hkust-gz.edu.cn}
\affiliation{Thrust of Advanced Materials \& Quantum Science and Technology Center, The Hong Kong University of Science and Technology (Guangzhou), 1 Duxue Rd., Nansha, Guangzhou, China}

\begin{abstract}
The recent discovery of long-sought dice flat band in layered YCl electride has opened up rich possibilities of correlation and topological physics in dice lattice systems~\cite{geng2026experimental, zhong2025intrinsic}. Here, we reveal a plethora of distinctive correlated topological phases in a generic layered dice lattice system at band filling of $\nu =4$ under on-site Hubbard interactions: (i) the system is an intrinsic sublattice anti-ferromagnetic (AFM) quantum spin-valley Hall insulator; (ii) a circularly polarized light (CPL) drives the AFM spin-valley insulator into a Floquet odd-parity $f$-wave altermagnet(AM) insulator; (iii) a vertical displacement field turns the Floquet $f$-wave AM insulator into a spin-valley-layer-polarized Chern insulator, with the sign of spin, valley and Chern number all controlled by the direction of the displacement field. Our results not only establish the layered dice lattice as a versatile platform for electrically switchable magnetic and topological phases, but also provide an all-electrical scheme for integrated spin-valley-layertronics for non-volatile information storage and processing.
\end{abstract}

\maketitle
\section{INTRODUCTION}

The dice lattice has been widely studied as a paradigmatic two-dimensional lattice framework for flat-band and topological phenomena, while theoretical studies have so far focused on coplanar arrangement of the hub and rim sites~\cite{soni2020flat, mohanta2023majorana, mondal2023topological}. A natural extension is to place different sublattice sites in separate layers, adding an extra layer degree of freedom that couples to existing spin and valley degrees of freedom~\cite{wang2011nearly}. The recent discovery of dice-lattice electronic bands in layered electride YCl exemplifies the physical relevance of this extension~\cite{geng2026experimental, zhong2025intrinsic}. In YCl, the effective dice lattice arises from interstitial anionic electrons rather than localized orbitals on coplanar atomic framework, with the anionic electrons on different sublattice sites confined to separate layers. This unique effective lattice architecture therefore motivates the exploration of unconventional magnetic states in which spin, valley, and layer degrees of freedom are all intrinsically intertwined.

\begin{figure}[t]
	\centering
	\includegraphics[width=0.48\textwidth]{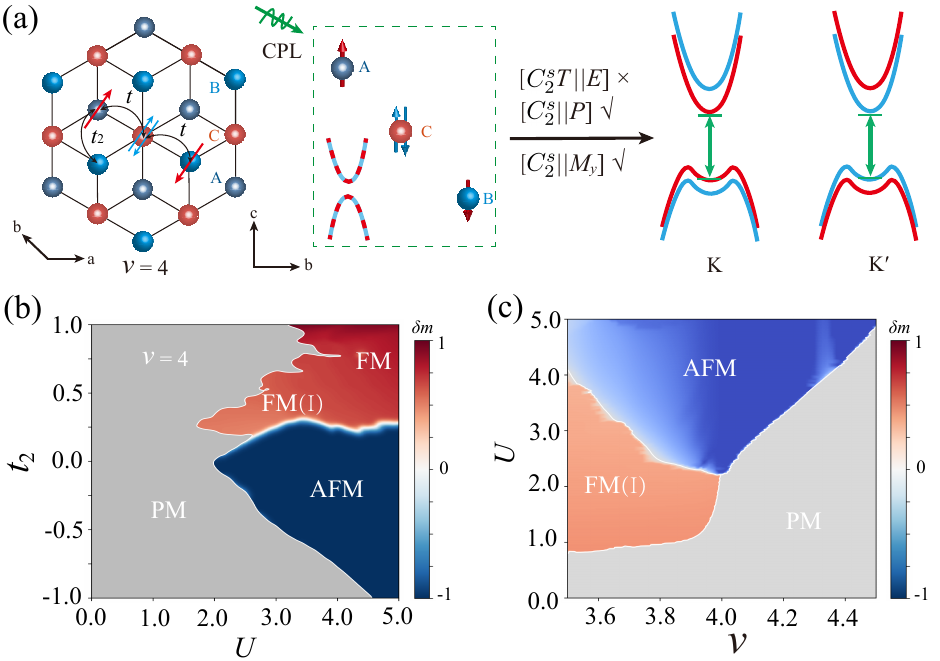}
	\caption{(a) Schematic dice-lattice geometry showing the antiferromagnetic (AFM) configuration at $\nu=4$ and the light-induced altermagnetic phase. Magnetic phase diagrams of the Hubbard model in Eq.~\eqref{Eq1} as functions of (b) $t_2$ and $U$, and (c) $U$ and $\nu$. In (b), $\epsilon_C=-0.3$ and $\nu=4$; in (c), $t_2=-0.1$ and $\epsilon_C=-0.3$.
	}  
	\label{fig1}
\end{figure}

Recent advances in the understanding of altermagnetism (AM), a distinctive form of compensated magnetism with zero net moment but finite momentum-dependent spin splitting~\cite{vsmejkal2022beyond, vsmejkal2022emerging, vsmejkal2020crystal, wu2007fermi, bai2024altermagnetism, zhang2025crystal, che2025engineering, pan2024general, banerjee2024altermagnetic, mondal2025distinguishing, ghorashi2024altermagnetic, reimers2024direct, ding2024large, yang2025three, li2025topological, karube2022observation, fedchenko2024observation, vsmejkal2023chiral, liu2024chiral, antonenko2025mirror, bhowal2024ferroically, zhu2025design, noh2025tunneling, jin2024skyrmion, ghorashi2025dynamical, xu2025alterpiezoresponse, wu2025intra, neehus2025projectively,huang2026light, zhu2026floquet, li2026floquet}, offers useful insights into how spin, valley and layer degrees of freedom can be integrated on a layered dice lattice. In addition to conventional even-parity class of $d$-, $g$-, or $i$-wave~\cite{feng2022anomalous, gonzalez2021efficient, shao2021spin, vsmejkal2022giant, karube2022observation}, odd-parity magnetism ($p$- or $f$-wave) obeys $E_{\mathbf{k},s}=E_{-\mathbf{k},-s}$, where spin splitting has opposite signs for time-reversed momenta. This odd-parity spin-momentum locking therefore offers a particularly appealing platform for spin and valley-related responses in the absence of spin-orbit coupling~\cite{lin2025odd, liu2026nonrelativistic, hu2025catalog, luo2026spin, zeng2026odd, hellenes2023p, brekke2024minimal}. In particular, altermagnetic states can arise spontaneously in bipartite lattices such as the Lieb lattice, where inequivalent sublattice environments allow magnetic order to generate altermagnetic phase~\cite{durrnagel2025altermagnetic, kaushal2025altermagnetism, ying2026magnons}. The dice lattice is likewise bipartite in nature, and its hexagonal symmetry is compatible with $f$-wave magnetism. Given the layered lattice architecture, should $f$-wave magnetism develops under interactions, the coupled spin-valley-layer physics is expected to emerge. The simplest Hubbard model with on-site Coulomb repulsion on a pristine dice lattice, however, supports robust antiferromagnetic order that respects the $\mathcal{PT}$ symmetry (Fig.\ref{fig1}b,c). This enforces spin degeneracy at each momentum point as in a conventional AFM which precludes the emergence of $f$-wave altermagnetism. 

Here, we predict that a circularly polarized light breaks the $\mathcal{PT}$-protection of the antiferromagnetic dice lattice and turns its spin-degenerate parent state into an $f$-wave topological altermagnet. In particuluar, spin-valley-layer locking (SVLL) occurs naturally under periodic driving of light fields, with opposite spin-valley sectors acquire distinct layer polarizations and topological masses. Importantly, a vertical electric field can selectively drive a band inversion in one particular spin-valley-layer sector. Reversing the gate polarity therefore switches the active topologically spin sector, realizing an electrically reconfigurable spin Chern filter. The same SVLL mechanism also enables a gate-switchable valley-selective optical transitions, which establishes a polarity-dependent circular-polarization selection rule. Building on the well-established coupled spin-valleytronic concepts developed in graphene and transition-metal dichalcogenide systems~\cite{ju2015topological, wu2019intrinsic, hung2019direct, zhou2019spin}, the spin-valley-layer polarization characterized with nontrivial topology in layered dice lattice offers a route to topological spin-valley-layertronics with electrically programmable and non-volatile information processing.

\section{MAGNETIC PHASE DIAGRAM of the dice lattice AT $\nu=4$}
We first examine the magnetic ground state of the quasi-three-dimensional dice lattice using the Hubbard model
\begin{equation} \label{Eq1}
\begin{aligned}
H=&-\sum_{\langle i, j\rangle, \sigma} t_{ij}c_{i\sigma}^{\dagger} c_{j\sigma}
+\epsilon_{C}\sum_{i \in C, \sigma} n_{i\sigma} 
+U\sum_i n_{i\uparrow} n_{i\downarrow},
\end{aligned}
\end{equation}
Here, $c_{i\sigma}^{\dagger}$ creates an electron with spin $\sigma$ on site $i$, and
$n_{i\sigma}=c_{i\sigma}^{\dagger}c_{i\sigma}$ is the corresponding number operator.
The hopping amplitude is taken to be $t_{AC/BC}=t$ on the $A$-$C$ and $B$-$C$ bonds, while $t_{AB}=t_2$ denotes the hopping between the two rim sites $A$ and $B$. The parameter $\epsilon_C$ controls the onsite energy of the hub site $C$, and $U$ is the local Hubbard repulsion.

To identify the magnetic phases, we perform a self-consistent Hartree-Fock calculation and characterize the resulting magnetic order by the sublayer-resolved moments. In particular, we introduce the new magnetic character parameter $\delta m$ to distinguish antiferromagnetic (AFM), ferromagnetic (FM), and ferrimagnetic (FM(I)) configurations, as detailed in Appendix~A.
Fig.~\ref{fig1}b shows the \(U\)-\(t_2\) phase diagram at the even filling \(\nu=4\), while Fig.~\ref{fig1}c follows the evolution of the magnetic ground state upon changing the filling.  The main result is that the AFM phase is stabilized over a broad and physically relevant region around \(\nu=4\).

The physical origin of the robust AFM phase at $\nu=4$ can be understood from the strong-coupling charge configuration. At this filling, the energetically favorable configuration is characterized by an almost doubly occupied hub site (C-site) and two nearly half-filled rim sites (A-site and B-site).
The magnetic moment on the hub site is therefore largely quenched, while the two rim sites carry the dominant local moments. Since the rim sites are close to half filling, the virtual hopping processes between them favor an AFM alignment through 180°-Anderson superexchange. In the simplest limit, the direct $A$-$B$ hopping gives an exchange scale $J_{AB}\sim 4t_2^2/U$, while additional indirect processes mediated by the hub site further renormalize the effective exchange interaction. As a result, the AFM configuration gains more kinetic exchange energy than the other configurations and becomes the lowest-energy magnetic state over a broad parameter regime. This mechanism is analogous to the altermagnetic Mott insulating state discussed in the modified Lieb lattice Hubbard model~\cite{kaushal2025altermagnetism}.

In this way, the magnetic phase diagrams not only establish the stability of the AFM ground state at $\nu=4$, but also clarify its relation to nearby competing magnetic orders. We therefore take this AFM phase as the parent magnetic state for the subsequent discussions.

\section{INTRINSIC quantum SPIN-VALLEY Hall insulator in the layered dice lattice at $\nu=4$}
Since the AFM is the ground state for the simple dice lattice at $\nu=4$, the tight-binding Hamiltonian of this AFM dice Hamiltonian can be written as in the basis $\Psi_{\mathbf{k}}=\left(c_{A\uparrow,\mathbf{k}},c_{C\uparrow,\mathbf{k}}, c_{B\uparrow,\mathbf{k}}, c_{A\downarrow,\mathbf{k}}, c_{C\downarrow,\mathbf{k}}, c_{B\downarrow,\mathbf{k}} \right)^T $
\begin{equation} \label{Eq2}
    H_{\mathrm{AFM}}(\mathbf{k})
    =
    \begin{pmatrix}
        h_{\uparrow}(\mathbf{k}) & 0 \\
        0 & h_{\downarrow}(\mathbf{k})
    \end{pmatrix},
\end{equation}
where
\begin{equation}
    h_s(\mathbf{k})
    =
    \begin{pmatrix}
        sM & -t\gamma^*(\mathbf{k}) & -t_2\gamma(\mathbf{k}) \\
        -t\gamma(\mathbf{k}) & \epsilon_C & -t\gamma^*(\mathbf{k}) \\
        -t_2\gamma^*(\mathbf{k}) & -t\gamma(\mathbf{k}) & -sM
    \end{pmatrix},
\end{equation}
here, $\gamma_{\mathbf{k}}=\sum_i e^{i\boldsymbol{k \cdot \delta_i}}$ for the nearest-neighboring hopping vector $\delta$. $s=+1\ $ for the spin up channel and $s=-1$ for the spin down. 
Equivalently, the AFM exchange field can be written as $H_M=M\sigma_z\otimes\Lambda_z$, where $\sigma_z$ is the Pauli matrix in the spin space and $\Lambda_z=\mathrm{diag}(1,0,-1)$ distinguishes the two rim sublayers $A$ and $B$.

To expose the intrinsic spin-valley Hall insulator, we perform the low-energy effective Hamiltonian up to second order of momentum $\mathbf{q}$ around the K/K$'$ valley
\begin{equation} \label{Eq6}
    h_{s\chi}^{0}(\mathbf q)
    =
    d_{0,\chi}(\mathbf q)\tau_0
    +
    \mathbf d_{s\chi}(\mathbf q)\cdot\boldsymbol{\tau},
\end{equation}
with
\begin{align}
    \mathbf d_{s,\chi}
    &=
    \left(
    \chi v_2  q_x-\mu(q_x^2-q_y^2),
    \chi v_2  q_y+2\mu q_xq_y,
    \chi \lambda^0_{s,\chi}
    \right),
\end{align}
where we define $d_{0,\chi}=-\mu(q_x^2+q_y^2)$ and $\mu=v^2/\epsilon_C$, $v=\frac{\sqrt{3}a}{2}t$ and $v_2=\frac{\sqrt{3}a}{2}t_2$. Here $\tau_i$ are Pauli matrices in the layer subspace with the basis $P_K=(A,B)$ for $\chi=+1$ and $P_{K'}=(B,A)$ $\chi=-1$. In the absence of light, the SVLL mass is given by the AFM exchange mass $\lambda^0_{s\chi}=sM$. For each SVLL low-energy Hamiltonian, the Chern number is evaluated as~\cite{qi2006topological}
\begin{equation}
    \mathcal{C}_{s\chi}
    =
    -\frac{1}{4\pi}
    \int d q_x d q_y\,
    \frac{
    \mathbf d_{s\chi}\cdot
    (\partial_{q_x}\mathbf d_{s\chi}\times
    \partial_{q_y}\mathbf d_{s\chi})
    }
    {|\mathbf d_{s\chi}|^3},
    \label{eq:main_valley_chern}
\end{equation}
here the integral is performed over the compactified valley plane. The quadratic term determines the asymptotic winding of $\mathbf d_{s\chi}$ at large $|\mathbf q|$ limit (Appendix D), leading to $\mathcal C^0_{s\chi}=\operatorname{sgn}(\chi\lambda^0_{s\chi})=\operatorname{sgn}(\chi sM)$, which is qualitatively different from that of a purely linear Dirac cone. Therefore, the total charge Chern number vanishes, $\sum_{s,\chi}C^0_{s,\chi}=0$, and the valley-summed Chern number for each spin also vanishes. Nevertheless, within a fixed valley the two spin sectors carry opposite Chern contributions. Thus, the spin-valley Chern number can be defined as 
\begin{equation}
    C^0_{\chi} = \frac{1}{2}(C^0_{\uparrow,\chi }-C^0_{\downarrow,\chi })=\chi \operatorname{sgn}(M),
\end{equation}
This identifies the equilibrium AFM phase at $\nu=4$ as an intrinsic spin-valley Hall insulator. A possible experimental signature of this phase can be accessed through an AFM domain wall separating regions with opposite Néel vectors (M$\rightarrow$-M)~\cite{ju2015topological}. Since such a reversal switches the SVLL Chern numbers ($C^0_{\chi}$), the domain wall is expected to host one-dimensional spin-valley-polarized conducting channels, with intervalley backscattering further suppressed for a sufficiently smooth interface. This scenario closely parallels the observation of topological valley transport along AB–BA domain walls in bilayer graphene~\cite{ju2015topological}, and provides a feasible route to probe the spin-valley topology in a layered-dice lattice.


\section{LIGHT-INDUCED f-WAVE ALTERMAGNETISM IN AFM DICE LATTICE}

Before introducing the optical field, we use the spin-group notation $[g_s\Vert g_l]$ to capture the symmetries of AFM layered-dice lattice, where $g_s$ and $g_l$ act in spin and lattice spaces, respectively. The unitary generators of Eq.~\ref{Eq2} AFM Hamiltonian have $[E\Vert C_{3z}], [C_2^s\Vert P], [C_2^s\Vert M_y]$, where $C_2^s$ is a $\pi$ spin rotation perpendicular to the ordered moment.  The parent AFM also has the antiunitary combined symmetry $[T\Vert P]\equiv [C_2^s\Vert P][C_2^sT\Vert E]$, which is the usual $PT$ symmetry in spin-group notation and protects the spin degeneracy of the equilibrium AFM bands.

\begin{figure}[t]
	\centering
	\includegraphics[width=0.48\textwidth]{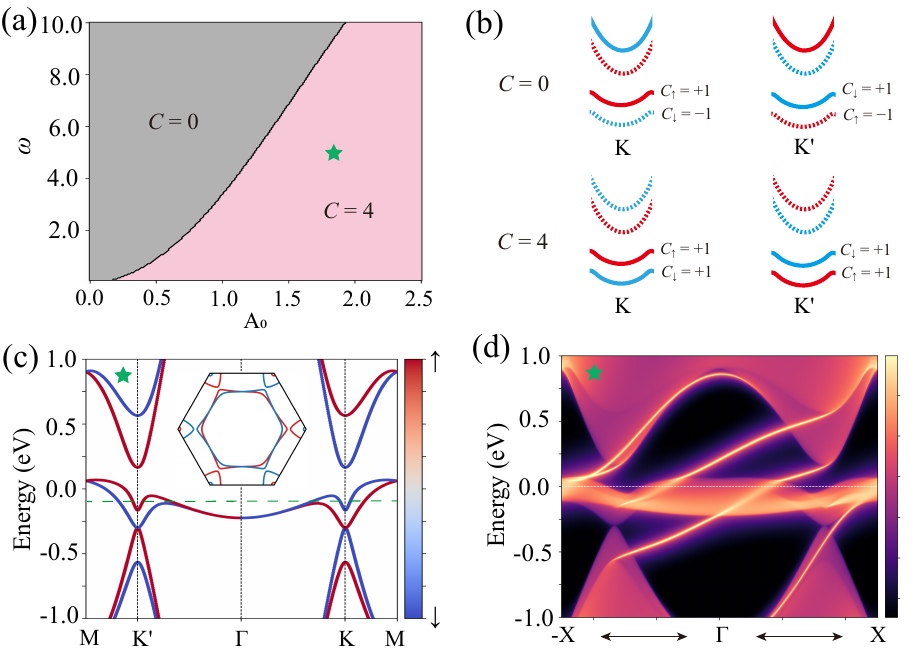}
	\caption{(a) The topological phase diagram as a function of light intensity and frequency. (b) The schematic of topological evolutions for the low and high CPL intensity at K/K$'$ valley. (c) The $f$-wave altermagnetic band structure for the green star labeled as (a). The inset shows the Fermi surface at $E_{F}=-0.12$ eV. (d) The edge states calculated for the light amplitude of $A_0=1.8$ \AA$^{-1}$ and $\omega=5.0$ eV. Parameters: $t=-1$ eV, $t_2=-0.1$ eV, $\epsilon=-0.3$ eV, $M=0.2$ eV.
	}  
	\label{fig2}
\end{figure}

Now, we introduce normally incident circularly polarized light (CPL) to construct the AM state by breaking the time-reversal symmetry in our case, which is introduced through the time-dependent vector potential $\mathbf{A}_{\eta}(t)=A_0(\eta\sin\omega t,\cos\omega t) $ 
where $\eta=\pm1$ denotes the different handed-CPL helicities. 
The coupling to electrons is included by the Peierls substitution
$\mathbf{k}\rightarrow\mathbf{k}+\mathbf{A}_{\eta}(t)$. Since the driven Hamiltonian is periodic in time, we expand it as
$H(\mathbf{k},t)=\sum_{n}H_n(\mathbf{k})e^{in\omega t}$. In the off-resonant high-frequency regime, the static effective Hamiltonian is obtained~\cite{kitagawa2011transport, bukov2015universal, eckardt2015high},
\begin{equation}
    H_{\mathrm{eff}}(\mathbf{k})
    =
    H_0(\mathbf{k})
    +
    \sum_{n\geq1}
    \frac{[H_{-n}(\mathbf{k}),H_n(\mathbf{k})]}{n\omega}
    +
    O(\omega^{-2}),
    \label{eq:heff_main}
\end{equation}
Keeping the leading one-photon process, the light-dressed Hamiltonian takes the form (more details in Appendix B)
\begin{equation} \label{Eq5}
    H_{\mathrm{eff}}(\mathbf{k})
    =
    H_{\mathrm{AFM}}^{(0)}(\mathbf{k})
    +
    \frac{\Delta_{\eta}(\mathbf{k})}{\omega}
    \sigma_0\otimes\Lambda_z ,
\end{equation}
where $H_{\mathrm{AFM}}^{(0)}$ denotes the static Hamiltonian with the hopping amplitudes renormalized by the zeroth-order Bessel function $J_0(\alpha A_0)$, where $\alpha=\frac{a}{\sqrt{3}}$ and $a$ is the lattice constant. The light-induced mass is $ \Delta_{\eta}(\mathbf{k})=4\sqrt{3}\eta (t^2-t_2^2) J^2_{1}(\alpha A_0) \sin(\frac{a k_x}{2})\left[\cos (\frac{\sqrt{3}ak_y}{2})-\cos (\frac{ak_x}{2})\right]$.
The sign of $\Delta_{\eta}$ is reversed by changing the light helicity, providing a direct optical knob for the band topology.  Physically, the Floquet massive term acts as a momentum-dependent sublayer-staggered mass, which converts the parent AFM order into an $f$-wave altermagnetic state, as described by Eq.~\ref{Eq5}. For a fixed helicity, CPL breaks the spin-preserving pseudo-time-reversal symmetry $[C_2^sT\Vert E]$, but preserves $[C_2^s\Vert P]$ by contrast (more details in Appendix B). 
The latter constrains the quasienergy spectrum according to $E_{s,n}(\mathbf{k})=E_{-s,n}(-\mathbf{k})$, implying an odd-parity spin splitting and opposite spin polarization at the $K$ and $K'$ valleys. Meanwhile, the residual symmetry $[C_2^s\Vert M_y]$ imposes $E_{s,n}(\mathbf{k})=E_{-s,n}(M_y\mathbf{k})$. Consequently, spin degeneracy is retained at momenta invariant under $M_y$, including the corresponding $\Gamma$--$M$ high-symmetry lines and their invariant points. Together with the $[E\Vert C_{3z}]$ symmetry, these symmetry-enforced nodal directions constrain the CPL-induced spin splitting to exhibit the characteristic $f$-wave altermagnetic feature shown in Fig.~\ref{fig2} c. 

We further investigate the layer-resolved band structure of the $f$-wave altermagnetic state in Fig.~\ref{figS2} a of Appendix E, which clearly reveals the spin-valley-layer locking (SVLL) features. Importantly, we examine the robustness of the altermagnetic order over a broad range of light intensities and Coulomb interactions, as shown in Fig.~\ref{figS1} of Appendix C. For appropriate interaction strengths, the AM order persists even with increasing light intensity, demonstrating its robustness against the Floquet driving.

\section{Light-driving topological phase transition}
In this section, we emphasize that CPL irradiation provides a route to engineer a high-Chern-number topological phase in the effective Floquet Hamiltonian. The optical field selectively modifies the SVLL mass terms and produces the redistribution of Berry curvature in momentum space. In the driven case, the low-energy Hamiltonian retains the form of Eq.~\ref{Eq6}, but with renormalized velocities $v=\frac{\sqrt{3}a}{2}tJ_0(\alpha A_0)$ and $v_2=\frac{\sqrt{3}a}{2}t_2J_0(\alpha A_0)$. More importantly, CPL introduces a valley-odd Floquet contribution to the static AFM mass, so that the SVLL mass becomes (Appendix D)
\begin{equation}
    \lambda_{s\chi}
    =
    sM+
    \chi F_\eta,\quad
    F_\eta=
    \frac{9\eta(t^2-t_2^2)J_1^2(\alpha A_0)}{\omega}.
    \label{Eq8}
\end{equation}
Thus the gap closing condition is satisfied as $\lambda_{s\chi}=0$, which means $|F_\eta|=|M|$. The scalar term proportional to $\tau_0$ does not affect the Berry curvature, while the quadratic off-diagonal term inherited from the static model fixes the valley winding. Using Eq.~\ref{eq:main_valley_chern}, the sector Chern number becomes $\mathcal C_{s\chi}=\operatorname{sgn}(\chi\lambda_{s\chi})$. In the low-light-intensity regime $|F_\eta|<|M|$, the Floquet term is not strong enough to change the sign pattern of the AFM masses. The system therefore remains a zero-charge-Chern spin-valley Hall insulator continuously connected to the equilibrium parent phase, as shown in Fig.~\ref{fig2} b. When the light intensity is increased such that $|F_\eta|>|M|$, the Floquet mass reverses selected SVLL sectors and converts the hidden spin-valley topology into a nonzero spin-resolved Chern number,
\begin{align}
\mathcal C_s=\sum_{\chi=\pm1}\mathcal C_{s\chi}
=
2\operatorname{sgn}(F_\eta),
\label{eq}
\end{align}
Consequently, for $t^2>t_2^2$, the total Chern number becomes $\mathcal C=4\eta$, which is dominated by the CPL-helicity. The full SVLL effective $k \cdot p$ model gives $\mathcal C_\uparrow=\mathcal C_\downarrow=2$ in this regime, consistent with the topological phase diagram in Fig.~\ref{fig2} a. The layer-resolved Berry curvature in Appendix E shows equivalent contributions from the two rim sublayers. The chiral edge states appearing for $|F_\eta|>|M|$, as shown in Fig.~\ref{fig2} d, further confirm the bulk-boundary correspondence of this Chern phase.

\section{Electrically-switchable Spin-Valley-layer Locking Physics}

\begin{figure}[t]
	\centering
	\includegraphics[width=0.48\textwidth]{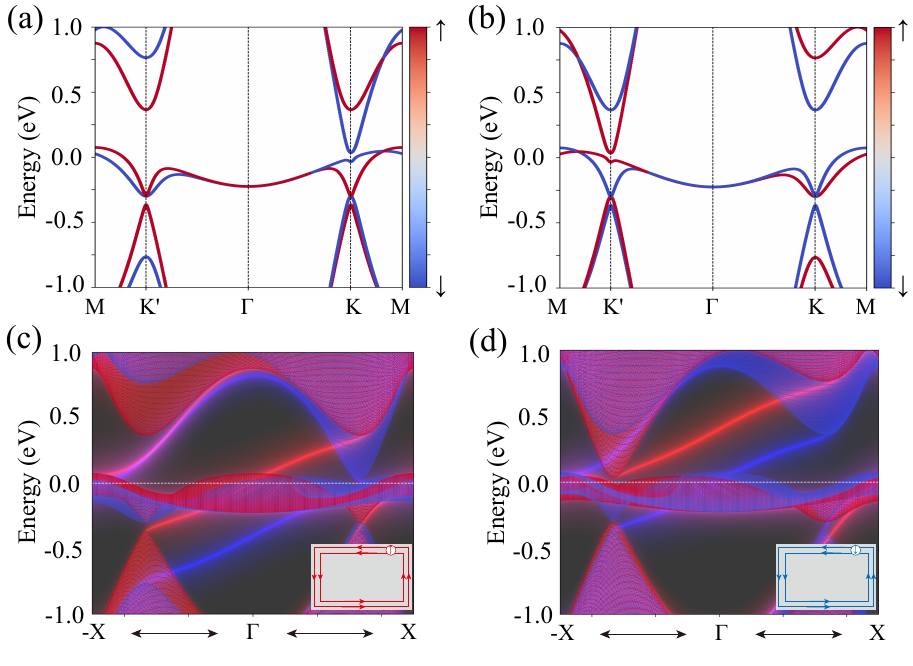}
	\caption{ The Band structures under an out-of-plane electric field along the $+z$ direction with $\delta<0$ (a) and along the $-z$ direction with $\delta>0$ (b). Corresponding electrically driven SVLL-selective edge spectra for $\delta<0$ (c) and $\delta>0$ (d), where only the spin-up and spin-down sectors host two chiral edge states, respectively. Parameters: $t=-1$ eV, $t_2=-0.1$ eV, $\epsilon=-0.3$ eV, $M=0.2$ eV, $A_0=1.8$ \AA$^{-1}$, $|\delta|=0.2$ and $\omega=5.0$ eV.
	}  
	\label{fig3}
\end{figure}

\subsection{Gating-Tunable Chern Phase}
We now use an electric field to switch the SVLL Chern phase, which can realize a spin-selective Chern filter: one spin sector remains topological while the other becomes trivial. To capture this effect, we introduce a layer-staggered gate potential into Eq.~\ref{Eq5},
\begin{equation}
    H_{\mathrm{E}}=\delta \sigma_0\otimes\Lambda_z, 
    \label{eq:gate_potential}
\end{equation}
where $\delta$ denotes the electrostatic potential difference between the two rim layers, represented by the $A$ and $B$ sublayers. The sign of $\delta$ specifies the gate polarity-$\delta>0$ corresponds to an electric field along the -$z$ direction, while $\delta<0$ corresponds to an electric field along the +$z$ direction. Since the gate term has the same layer matrix as the AFM and Floquet masses, the valley mass in Eq.~\ref{Eq8} becomes $\lambda_{s\chi}(\delta)=sM+\chi F_\eta+\delta$. The SVLL Chern number is therefore $\mathcal C_{s\chi}(\delta)=\operatorname{sgn}\left[F_\eta+\chi(sM+\delta)\right]$. Summing over the two valleys gives the spin-resolved Chern number
\begin{equation}
    \mathcal C_s(\delta)
    =
    2\,\operatorname{sgn}(F_\eta)\,
    \Theta\!\left(|F_\eta|-|sM+\delta|\right),
    \label{eq:gated_spin_chern_window}
\end{equation}
where $\Theta(x)$ is the step function. Eq.~\ref{eq:gated_spin_chern_window} shows that the two spin sectors have topological windows centered at opposite gate biases, $\delta=-sM$. Therefore, sweeping the layer-staggered gate potential can selectively drive one spin sector across the valley band inversion while leaving the other spin sector topologically trivial. A spin-selective Chern filter is obtained for $F_\eta>M>0$ when
\begin{align}
    & \delta<0:\quad
    -F_\eta-M<\delta<-F_\eta+M,
    \quad
    & \mathcal C_\uparrow=2,\ \mathcal C_\downarrow=0,
    \nonumber\\
    & \delta>0:\quad
    F_\eta-M<\delta<F_\eta+M,
    \quad
    & \mathcal C_\downarrow=2,\ \mathcal C_\uparrow=0,
    \label{eq:gate_windows}
\end{align}
Note that a fixed nonzero gate bias breaks the $[C_2^s\Vert P]$ and $[C_2^s||M_y]$ symmetry, since inversion exchanges the two rim layers and changes $\delta\Lambda_z$ into $-\delta\Lambda_z$. The band structures for opposite out-of-plane electric fields are presented in Fig.~\ref{fig3} a and \ref{fig3} b. When the electric field is applied along the $+z$ direction, the band inversion is first triggered in the spin-down sector at the $K$ valley, while the spin-down bands at $K'$ remain uninverted. Consequently, the Berry-curvature contributions from the two valleys cancel within the spin-down sector, yielding a vanishing spin-resolved Chern number. At the same time, the inverted spin-down $K$ sector develops a Berry-curvature distribution opposite to that of the spin-up $K'$ sector, reflecting the characteristic SVLL correspondence. Reversing the electric field to the $-z$ direction switches the inversion to the spin-up $K'$ valley, while its $K$ counterpart remains uninverted. Therefore, the direction of the electric field provides an efficient means to selectively control the SVLL band inversion and the associated Berry-curvature distribution. The corresponding topological features are corroborated by the edge states shown in Fig.~\ref{fig3} c and d.

\subsection{Gating-Controlled Optical selection}
The optical selection between the two valleys can be controlled by the external electric field for the highest valence band and the lowest conduction band. In the absence of an electric field, the Berry curvature distributions at the $K$ and $K'$ valleys are identical, protected by the $[C_2^s\Vert P]$ and $[C_2^s\Vert M_y]$ symmetries. As a result, the two valleys exhibit equivalent optical responses, and no SVLL-selective optical excitation is expected. Once an out-of-plane electric field is applied, these symmetries are broken. The electric field further drives SVLL-dependent band inversion, which reverses the sign of the Berry curvature in selected SVLL sectors. This symmetry breaking and Berry curvature reversal lift the equivalence between the two valleys and enable gate-controlled optical selection. The optical selectivity for CPL is characterized by the degree of polarization~\cite{li2026quantum}
\begin{equation}
\eta_C(\boldsymbol{k})=-\frac{\Omega_{v c}^{x y}(\boldsymbol{k})}{g_{v c}^{x x}(\boldsymbol{k})+g_{v c}^{y y}(\boldsymbol{k})},
\end{equation}
where $g_{vc}^{\alpha\beta}(\mathbf{k})$ and $\Omega_{vc}^{\alpha\beta}(\mathbf{k})$ represent the symmetric and antisymmetric components of the interband quantum geometric tensor, respectively. The indices $v$ and $c$ denote the lower and upper bands of the effective $k\cdot p$ model Eq.~\ref{Eq6}, mainly contributed by $(s,\chi)=(+1,-1)$ and $(-1,+1)$, within which $\Omega^{xy}_{vc}$ and $g^{ij}_{vc}$ are evaluated. 

\begin{figure}[t]
	\centering
	\includegraphics[width=0.48\textwidth]{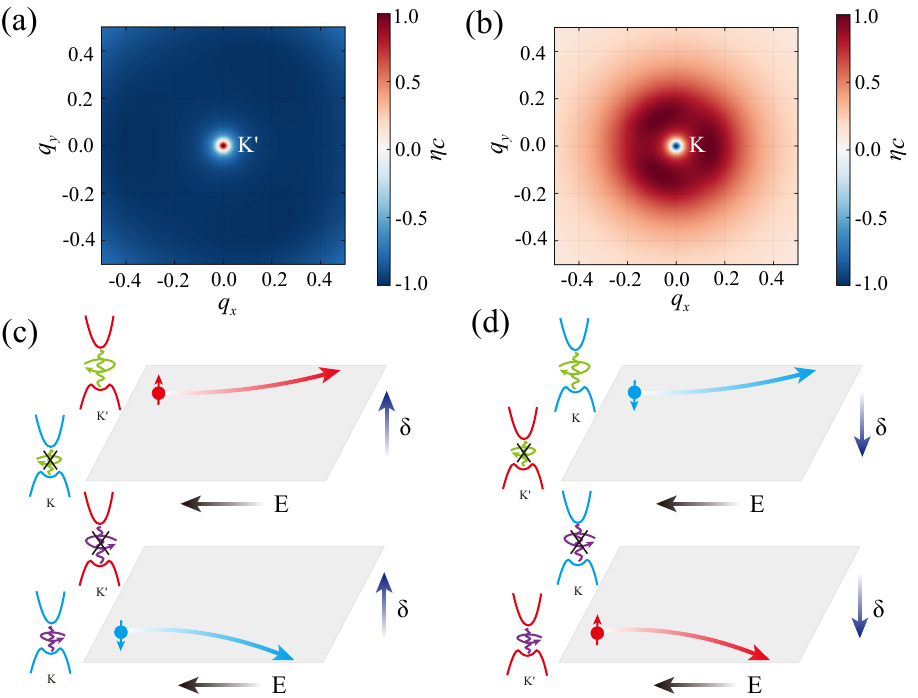}
	\caption{ (a),(b) Calculated the SVLL-dependent degree of circular polarization $\eta_C$ near the $K'$ and $K$ valleys under a $+z$ direction electric field with $\delta<0$, respectively. (c),(d) Schematic of the gate-controlled SVLL optical selection and spin deflection. Under one gate polarity, the $K'$ and $K$ valleys are selected by opposite helical CPL. The photoexcited electrons show the same longitudinal current but opposite transverse anomalous currents for the distinct CPL. Reversing the electric-field direction reverses the Berry-curvature sign of the gate-selected SVLL sector, switches the spin polarization, and interchanges the valley selected by a given light helicity. The parameters are consistent with those shown in Fig.~\ref{fig3}.
    }
	\label{fig4}
\end{figure}

When we apply the $+z$ direction electric field, corresponding to $\delta<0$, the Berry curvature of the spin-down $K$ valley changes sign first at $\delta=-(F_\eta-M)$. For the two SVLL sectors relevant to the optical transition, the degree of circular polarization is mainly governed by the sign of the Berry curvature near the valley center
\begin{align}
    \eta_C^{\uparrow K'}
    &=
    \operatorname{sgn}
    \left(
    F_\eta-M-\delta
    \right),
    \nonumber\\
    \eta_C^{\downarrow K}
    &=
    \operatorname{sgn}
    \left(
    F_\eta-M+\delta
    \right).
\end{align}
Therefore, a negative gate bias reverses the Berry curvature and circular selectivity of the spin-down $K$ valley, while leaving the spin-up $K'$ valley unchanged. As shown in Fig.~\ref{fig4} a and b, the calculated degree of circular polarization takes opposite signs at the $K'$ and $K$ valleys. This behavior differs fundamentally from conventional spin-valley locking, where the valley-selective optical rule is imposed by the intrinsic symmetry of the system~\cite{xiao2012coupled, yao2008valley}. Here, the two valleys are optically equivalent without an electric field. The valley contrast appears only when the out-of-plane electric field breaks the $[C_2^s\Vert P]$ and $[C_2^s\Vert M_y]$ symmetries and induces an SVLL-dependent reversal of the Berry curvature, thereby enabling electrically switchable optical selection. Conversely, a positive gate bias reverses the Berry curvature of the spin-up $K'$ valley and switches its preferred circular polarization. 

Fig.~\ref{fig4} c and d schematically illustrate the electrically switchable optical selection. For a fixed gate polarity, the optical transition becomes helicity selective in each SVLL sector. Left-handed CPL predominantly excites the electrons of the spin-up $K'$ valley, for example, while the spin-down $K$ valley is optically inactive for the same helicity. Under an in-plane electric field, the photoexcited spin-up carriers in the $K'$ valley are deflected to one transverse side. Conversely, right-handed CPL selectively excites the spin-down $K$ valley, whose anomalous deflection occurs in the opposite transverse direction. Reversing the out-of-plane electric field switches the active SVLL sectors, thereby interchanging the valley and spin polarization selected by a given light helicity. The holes carry opposite spins and propagate in opposite directions. This electrically switchable SVLL and valley-contrasting optical selection provide a tunable route toward gate-controllable spintronic and valleytronic functionalities.


\section{CONCLUSIONS and outlooks}

In this work, we propose a Floquet and gate-controlled route to engineer and control high-Chern-number phases in the altermagnetic dice lattice. We first investigate an interacting dice lattice at filling number $\nu=4$ and identify a collinear antiferromagnetic ground state as the parent phase for light-induced altermagnetism. By deriving the equilibrium valley $k\cdot p$ model, we show that this AFM state is an intrinsic quantum spin-valley Hall insulator - the total charge Chern number is zero, while opposite spin sectors within a fixed valley carry a nonzero spin-valley Chern number. We further show that circularly polarized light generates a spin-valley-layer locking (SVLL)-dependent Floquet mass, drives the system into a high-Chern-number ($C=\pm4$) altermagnetic phase, and produces chiral edge states. We then demonstrate that a layer-staggered gate potential can selectively reverse the Berry curvature in a chosen SVLL sector, realizing an electrically switchable spin Chern filter and valley-contrasting optical selection. These findings identify SVLL as a unified mechanism for converting electrical gating into spin-selective topology and valley-dependent optical functionality, suggesting a pathway toward multifunctional devices in which spin, valley, and layer information can be encoded and manipulated without relying on macroscopic magnetization.

In the dice lattice, the flat band is determined by the phase configuration of the compact localized Wannier functions, whose amplitudes on the $A$ and $B$ rim sublattices have opposite phases and therefore produce exactly cancelling hopping contributions at each $C$ hub site. The experimental realization therefore hinges on constructing the characteristic geometry and connectivity of the dice lattice, which provide the structural basis for the phase cancellation responsible for the flat band, instead of orbital components in a specific chemical environment. Potential proposals to realize the dice geometry include SrTiO$_3$/SrIrO$_3$/SrTiO$_3$ trilayer heterostructure grown in the (111) direction~\cite{okamoto2018transition} and the LaAlO$_3$/SrTiO$_3$ (111) quantum wells~\cite{doennig2013massive}. However, these proposals rely on specific crystallographic orientations and cleavage planes, which may limit their practical application. By contrast, alternative routes to realizing dice lattices have been proposed, including molecular manipulation of CO molecules on Cu(111)~\cite{tassi2024implementation} and the emergence of an effective dice lattice from the intrinsic charge-density distribution in pristine YCl~\cite{geng2026experimental}. These complementary realizations highlight the feasibility of engineering or identifying dice-lattice electronic structures in realistic systems and may stimulate the search for intrinsic material platforms hosting layer-dependent dice lattices and their associated magnetic and topological phenomena.


\section{ACKNOWLEDGMENTS}
J. Zhong acknowledges Yongheng Ge and Xiaoxu Wang for helpful discussions. B.T. Zhou acknowledges the support of NSFC-Young Scientists Fund (No.12504194), Guangdong Provincial Quantum Science Strategic Initiative (No. GDZX2501004), Guangdong Provincial Talents Program (No. 2025D03J0006) and Start-up Fund of HKUST(GZ) (No. G0104000263). J.Y. Zou acknowledges the support of National Natural Science Foundation of China (No. 12404182).

\appendix
\section{The definition of magnetic character}

In this work, the magnetic state is characterized directly from the sublayer-resolved spin polarization
\begin{equation}
    \mathbf{S}=(S_A,S_B,S_C),
\end{equation}
where \(S_\alpha=(n_{\alpha\uparrow}-n_{\alpha\downarrow})/2\) is the local spin moment on sublayer \(\alpha=A,B,C\).  Instead of assigning a phase only by comparing the signs of the three moments, we decompose
\(\mathbf{S}\) into three orthonormal magnetic channels,
\begin{align}
    m_F &=
    \left| \frac{S_A+S_B+S_C}{\sqrt{3}} \right|, \\
    m_{AF} &=
    \left| \frac{S_A-S_B}{\sqrt{2}} \right|, \\
    m_X &=
    \left| \frac{S_A-2S_C+S_B}{\sqrt{6}} \right|,
\end{align}
Here \(m_F\) measures the uniform FM component, while \(m_{AF}\) measures the staggered component between the two outer sublayers \(A\) and \(B\).  The third component \(m_X\) is orthogonal to both of these channels and measures the imbalance of the middle sublayer moment relative to the two outer sublayers, which is therefore useful for distinguishing a FM(I) character from a purely FM one.

The magnetic character shown in the phase diagram of main text is quantified by
\begin{equation}
    \delta m =
    \frac{m_F-m_{AF}}
    {m_F+m_{AF}+m_X+\epsilon},
    \label{eq:Cstar}
\end{equation}
where \(\epsilon\) is a small positive number introduced only to avoid an ill-defined ratio when the total magnetic moment is numerically zero. With this convention, \(\delta m>0\) indicates a predominantly FM-like character, whereas \(\delta m<0\) indicates a predominantly AFM-like character.  The inclusion of \(m_X\) in the denominator prevents FM(I) configurations from being artificially collapsed onto the fully FM limit.

In regions where the total magnetic moment
\begin{equation}
    |\mathbf{S}|=\sqrt{S_A^2+S_B^2+S_C^2}.
\end{equation}
is below the numerical resolution of the calculation, we do not assign a magnetic phase.  These regions are shown as paramagnetic regions in the phase diagram.  Likewise, small isolated FM-like features with moments of order \(10^{-3}\) are treated as numerical uncertainty rather than as stable magnetic phases.

\section{Derivation of the Floquet mass}

This Appendix derives the light-induced effective Floquet Hamiltonian quoted in Eq.~\eqref{Eq5}. We use the basis $\Psi_{\mathbf{k}}=\left(c_{A\uparrow,\mathbf{k}}, c_{C\uparrow,\mathbf{k}}, c_{B\uparrow,\mathbf{k}}, c_{A\downarrow,\mathbf{k}},c_{C\downarrow,\mathbf{k}}, c_{B\downarrow,\mathbf{k}}\right)^T$ and choose the three nearest-neighboring vectors as $\boldsymbol{\delta}_1=\alpha(0,1), \boldsymbol{\delta}_2=\alpha(-\frac{\sqrt{3}}{2},-\frac{1}{2}), \boldsymbol{\delta}_3=\alpha(\frac{\sqrt{3}}{2},-\frac{1}{2})$, where $\alpha=a/\sqrt{3}$.  We write $k_j=\mathbf{k}\cdot\boldsymbol{\delta}_j$ and $\gamma(\mathbf{k})=\sum_{j=1}^3e^{ik_j}$.
For a normally incident circularly polarized field $\mathbf{A}_{\eta}(t)=A_0(\eta\sin\omega t,\cos\omega t)$, where $\eta=\pm1$ represents the left-handed CPL and right-handed CPL, respectively. The Peierls substitution gives
$e^{ik_j}\rightarrow e^{ik_j}e^{i\mathbf{A}_\eta(t)\cdot\boldsymbol{\delta}_j}$. It is useful to parametrize the bond-dependent phase by
\begin{equation}
    \mathbf{A}_\eta(t)\cdot\boldsymbol{\delta}_j
    =
    \alpha A_0\sin(\omega t+\phi_j^\eta),
    \quad
    e^{i\phi_j^\eta}
    =
    \eta\frac{\delta_{j,x}}{\alpha}
    +i\frac{\delta_{j,y}}{\alpha}.
    \label{eq:app_phi_def}
\end{equation}
With the convention $ H_n(\mathbf{k})= \frac{1}{T}\int_0^Tdt\,H[\mathbf{k}+\mathbf{A}_\eta(t)]e^{-in\omega t}$ with the driving frequency $\omega=2\pi/T$ and using the Jacobi-Anger expansion, one obtains 
\begin{align}
    \gamma_\eta^{(n)}(\mathbf{k})
    &=
    J_n(\alpha A_0)
    \sum_{j=1}^3 e^{ik_j}e^{in\phi_j^\eta},
    \label{eq:app_gamma_n}\\
    \bar{\gamma}_\eta^{(n)}(\mathbf{k})
    &\equiv
    \left[\gamma_\eta^{(-n)}(\mathbf{k})\right]^*
    =
    (-1)^nJ_n(\alpha A_0)
    \sum_{j=1}^3 e^{-ik_j}e^{in\phi_j^\eta}.
    \label{eq:app_gamma_bar_n}
\end{align}
Here $\bar{\gamma}_\eta^{(n)}$ denotes the Fourier coefficient of $\gamma^*(\mathbf{k}+\mathbf{A}_\eta(t))$. This notation is kept explicit to avoid confusing complex conjugation with the Fourier index.

The zeroth Fourier component simply renormalizes the hopping structure,
\begin{equation}
    \gamma_\eta^{(0)}(\mathbf{k})=J_0(\alpha A_0)\gamma(\mathbf{k}),
    \quad
    \bar{\gamma}_\eta^{(0)}(\mathbf{k})=J_0(\alpha A_0)\gamma^*(\mathbf{k}).
\end{equation}
For $n\neq0$, the static onsite and exchange terms do not contribute because $T^{-1}\int_0^Tdt\,e^{-in\omega t}=0$.  Thus each spin block has
\begin{equation}
    h^{(n)}(\mathbf{k})
    =
    -
    \begin{pmatrix}
        0 & t\bar{\gamma}_\eta^{(n)} & t_2\gamma_\eta^{(n)}\\
        t\gamma_\eta^{(n)} & 0 & t\bar{\gamma}_\eta^{(n)}\\
        t_2\bar{\gamma}_\eta^{(n)} & t\gamma_\eta^{(n)} & 0
    \end{pmatrix},
    \quad n\neq0 .
    \label{eq:app_hn}
\end{equation}

In the off-resonant regime, the leading van Vleck correction is
\begin{equation}
    H_{\mathrm{eff}}(\mathbf{k})
    =
    H_0(\mathbf{k})
    +
    \frac{[H_{-1}(\mathbf{k}),H_1(\mathbf{k})]}{\omega}
    +O(\omega^{-2}),
    \label{eq:app_vv}
\end{equation}
Since the drive does not mix spins, the commutator can be evaluated within a single $3\times3$ block.  Direct multiplication of Eq.~\eqref{eq:app_hn} gives
\begin{equation}
    [h^{(-1)},h^{(1)}]
    =
    \Delta_\eta(\mathbf{k})\Lambda_z,
    \quad
    \Lambda_z=\mathrm{diag}(1,0,-1).
    \label{eq:app_commutator}
\end{equation}
with
\begin{equation}
    \Delta_\eta(\mathbf{k})
    =
    (t^2-t_2^2)
    \left[
    \gamma_\eta^{(1)}\bar{\gamma}_\eta^{(-1)}
    -
    \gamma_\eta^{(-1)}\bar{\gamma}_\eta^{(1)}
    \right],
    \label{eq:app_delta_compact}
\end{equation}
Substituting Eqs.~\eqref{eq:app_gamma_n} and
\eqref{eq:app_gamma_bar_n} yields the bond-coordinate form
\begin{align}
    \Delta_\eta(\mathbf{k})
    &=
    4\sqrt{3}\,\eta\,(t^2-t_2^2)J_1^2(\alpha A_0)
    \sin\left(\frac{ak_x}{2}\right)
    \nonumber\\
    &\quad\times
    \left[
    \cos\left(\frac{\sqrt{3}ak_y}{2}\right)
    -
    \cos\left(\frac{ak_x}{2}\right)
    \right],
    \label{eq:app_delta_cartesian}
\end{align}

Equations~\eqref{eq:app_commutator}--\eqref{eq:app_delta_cartesian} therefore give the leading light-induced correction
\begin{equation}
    H_{\mathrm{F}}(\mathbf{k})
    =
    \frac{\Delta_\eta(\mathbf{k})}{\omega}\,
    \sigma_0\otimes\Lambda_z,
    \label{eq:app_HF}
\end{equation}

Combining the zeroth-order hopping renormalization with Eq.~\eqref{eq:app_HF}, the effective Hamiltonian can be written as $H_{\mathrm{eff}}(\mathbf{k})=H^0_{\mathrm{AFM}}(\mathbf{k})+H_{\mathrm{F}}(\mathbf{k})$, where 
\begin{equation}
    h^{\mathrm{eff}}_s(\mathbf{k})
    =
    \begin{pmatrix}
        sM
        & -t\bar{\gamma}_\eta^{(0)}
        & -t_2\gamma_\eta^{(0)}
        \\
        -t\gamma_\eta^{(0)}
        & \epsilon_C
        & -t\bar{\gamma}_\eta^{(0)}
        \\
        -t_2\bar{\gamma}_\eta^{(0)}
        & -t\gamma_\eta^{(0)}
        & -sM
    \end{pmatrix},
    \label{eq:app_heff_spin}
\end{equation}
here, $H^0_{\mathrm{AFM}}=\mathrm{diag}(h^{\mathrm{eff}}_\uparrow, h^{\mathrm{eff}}_\downarrow)$. To capture the physical meaning, we redefine $m_s(\mathbf{k})=sM+\Delta_\eta(\mathbf{k})/\omega$ and $s=+1$ ($-1$) labels the spin-up (spin-down) sector.

Here, we give the brief proving that light does indeed induce altermagnetism. The symmetry relation of the spin-resolved spectrum can be seen directly from the effective mass term. Since the Floquet mass is odd in momentum, $\Delta_\eta(-\mathbf{k})=-\Delta_\eta(\mathbf{k})$, one has 
\begin{equation} \label{eq:mass_relation_cp}
    m_{-s}(-\mathbf{k})=-m_s(\mathbf{k}),
\end{equation}
The spatial part of the spin-space-group operation $[C_2\|P]$ exchanges the $A$ and $B$ sublayers while reversing momentum. In the sublayer basis $(A,C,B)$, the inversion operator can be represented as
\begin{equation}
    U_P=
    \begin{pmatrix}
        0&0&1\\
        0&1&0\\
        1&0&0
    \end{pmatrix},
\end{equation}
Using $\gamma_\eta^{(0)}(-\mathbf{k})=\bar{\gamma}_\eta^{(0)}(\mathbf{k})$, Eq.~\eqref{eq:mass_relation_cp} gives
\begin{equation}
    U_P h_s^{\mathrm{eff}}(\mathbf{k})U_P^\dagger
    =
    h_{-s}^{\mathrm{eff}}(-\mathbf{k}),
    \label{eq:cp_relation_hamiltonian}
\end{equation}
Therefore the two Hamiltonians are unitarily equivalent, and their spectra satisfy
\begin{equation}
    E_{s,n}(\mathbf{k})=E_{-s,n}(-\mathbf{k}),
    \label{eq:cp_relation_energy}
\end{equation}

In contrast, for a fixed spin sector,
\begin{equation}
    m_s(-\mathbf{k})
    =
    sM-\frac{\Delta_\eta(\mathbf{k})}{\omega},
\end{equation}
which is generally different from $m_s(\mathbf{k})$. More explicitly,
\begin{equation}
    m_s^2(\mathbf{k})-m_s^2(-\mathbf{k})
    =
    4sM\frac{\Delta_\eta(\mathbf{k})}{\omega},
    \label{eq:mass_square_difference}
\end{equation}
Thus, when both the AFM exchange field and the light-induced Floquet mass are finite, namely $M\neq0$ and $\Delta_\eta(\mathbf{k})\neq0$, the single-spin spectrum is generally nonreciprocal,
\begin{equation}
    E_{s,n}(\mathbf{k})\neq E_{s,n}(-\mathbf{k}).
    \label{eq:single_spin_nonreciprocal}
\end{equation}
Hence the $[C_2\|P]$ symmetry enforces the relation between opposite spin sectors at opposite momenta, while it does not require each individual spin sector to be symmetric under $\mathbf{k}\rightarrow-\mathbf{k}$.


\section{Robustness of Altermagnetic order}
\begin{figure}[t]
	\centering
	\includegraphics[width=0.48\textwidth]{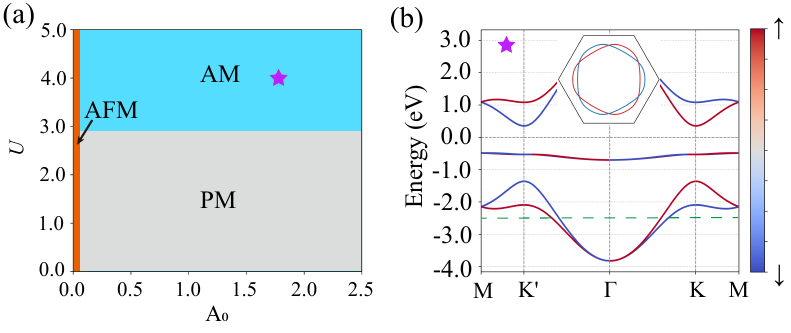}
	\caption{ (a) The magnetic phase diagram as a function of the light intensities and Coulomb interactions. (b) The HF band structure for the purple star point. The inset shows the Fermi surface at $E_F=-0.25$ eV labeled the green dash line.
	}  
	\label{figS1}
\end{figure}

We further examine the differences of magnetic order against variations in the light intensity and Coulomb interaction, the Floquet-Hubbard model can be wriiten as
\begin{equation}
    \begin{aligned}
    H_{\mathrm{F-H}} =     &\sum_{\mathbf{k}} \Psi^\dagger_{\mathbf{k}} \left[H_{\mathrm{AFM}}^{(0)}(\mathbf{k})|_{M=0}+\frac{\Delta_{\eta}(\mathbf{k})}{\omega} \sigma_0\otimes\Lambda_z\right]\Psi_{\mathbf{k}} \\
    &+U\sum_i n_{i\uparrow} n_{i\downarrow}.
    \end{aligned}
\end{equation}
where we consider the non-magnetic state ($M=0$) for the first term and the renormalizations by the period light. Fig.~\ref{figS1} a summarizes the magnetic phase diagram obtained from our Hartree–Fock calculations. In the absence of light irradiation, the system exhibits a $\mathcal{PT}$-symmetric AFM state. As the light intensity increases, the magnetic phase diagram reflects a competition between Floquet driving and Coulomb interactions. Within an appropriate range of interaction strengths, this competition stabilizes the AM phase. The corresponding HF band structure is shown in Fig.~\ref{figS1} b.

\section{Valley \texorpdfstring{$k\cdot p$}{k.p} model}

We derive the effective valley $k\cdot p$ model starting from the two inequivalent valleys
\begin{equation}
    \mathbf{K}_\chi=\chi\left(\frac{4\pi}{3a},0\right),
    \qquad \chi=\pm1 .
\end{equation}
Expanding around $\mathbf{k}=\mathbf{K}_\chi+\mathbf{q}$, the structure factor takes the linear form
\begin{equation}
	\Gamma_\chi(\mathbf q)\simeq
	J_0(\alpha A_0)\frac{\sqrt{3}a}{2}\left(-\chi q_x+i q_y\right),
\end{equation}
while the Floquet-induced valley mass is
\begin{equation}
    \Delta_\eta(\mathbf{K}_\chi)
    =
    9\eta\chi(t^2-t_2^2)J_1^2(\alpha A_0),
\end{equation}

So that $\lambda_{s,\chi}=sM+\Delta_{\eta}(\mathbf{K}_\chi)/\omega$. The onsite level $\epsilon_C$ is the remote low-lying subspace in the parameter regime used in the main text, and it does not participate in the valley band inversion. We therefore downfold this one-dimensional subspace out. Because the mirror operation $M_y$ relates the two valleys, we choose the active basis as
\begin{equation}
    P_{K}=(A,B),\qquad
    P_{K'}=(B,A),
    \label{eq:appC_valley_basis}
\end{equation}
where the order of the two active basis states is reversed between the two valleys. The remaining state with onsite energy $\epsilon_C$ is denoted by $Q$.

We perform the Löwdin downfolding in the basis $P_\chi\oplus Q$, yielding the effective two-band Hamiltonian at $K$ valley
\begin{align}
    h_{s,+}^{kp}(\mathbf q)
    &=
    \begin{pmatrix}
        \lambda_{s,+} & -t_2\Gamma_+\\
        -t_2\Gamma_+^* & -\lambda_{s,+}
    \end{pmatrix}
    -
    \frac{t^2}{\epsilon_C}
    \begin{pmatrix}
        |\Gamma_+|^2 & \Gamma_+^{*2}\\
        \Gamma_+^2 & |\Gamma_+|^2
    \end{pmatrix},
    \label{eq:appC_lowdin_K}
\end{align}
while at $K'$ valley it is
\begin{align}
    h_{s,-}^{kp}(\mathbf q)
    &=
    \begin{pmatrix}
        -\lambda_{s,-} & -t_2\Gamma_-^*\\
        -t_2\Gamma_- & \lambda_{s,-}
    \end{pmatrix}
    -
    \frac{t^2}{\epsilon_C}
    \begin{pmatrix}
        |\Gamma_-|^2 & \Gamma_-^2\\
        \Gamma_-^{*2} & |\Gamma_-|^2
    \end{pmatrix}.
    \label{eq:appC_lowdin_Kp}
\end{align}
So the leading quadratic Dirac terms give Eq.~\ref{Eq6} in the main text.

This quadratic term dominates the large-momentum behavior and fixes the winding of the in-plane vector $(d_x,d_y)$. Then the in-plane texture can be written as
$d_x+i d_y=\chi v_2 z-\mu \bar{z}^{,2}$, where $z=q_x+iq_y$. Using $z=re^{i\theta}$, one finds that the asymptotic texture is governed by $e^{-2i\theta}$, giving a winding number $N_\infty=-2$. As a result, the SVLL Chern number becomes $\mathcal C_{s\chi}=\operatorname{sgn}(\chi\lambda_{s\chi})$ in the main text.

\section{Layer polarization and layer-resolved Berry curvature}

\begin{figure}[t]
	\centering
	\includegraphics[width=0.48\textwidth]{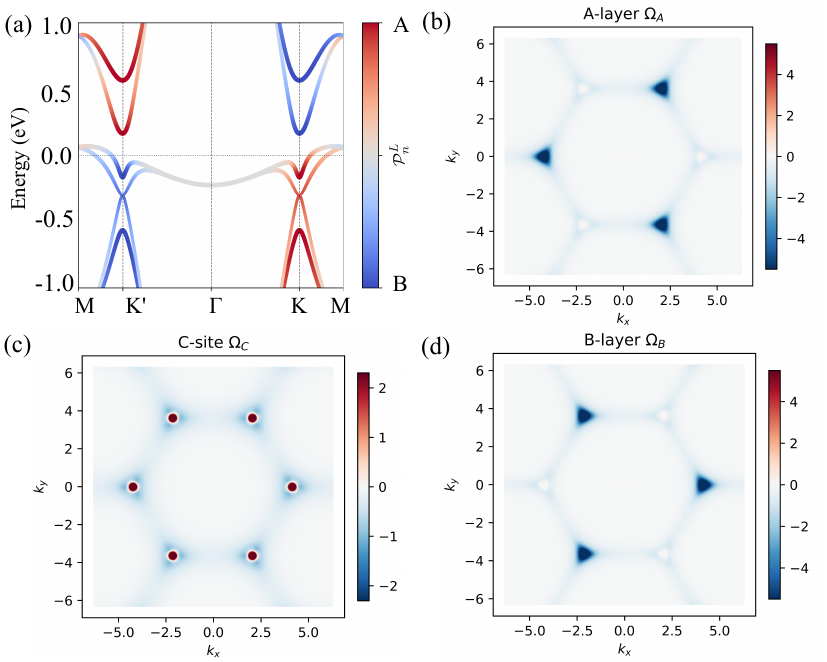}
	\caption{ (a) layer-resolved band structure for the rim sublayers. (b)-(d) layer-resolved Berry curvature. 
	}  
	\label{figS2}
\end{figure}

Here we describe the numerical diagnostics used to identify the spin-valley-layer locking. The layer degree of freedom is represented by the two rim sublayers $A$ and $B$. We define the projection operators
\begin{align}
    P_A&=\mathrm{diag}(1,0,0,1,0,0),\\
    P_B&=\mathrm{diag}(0,0,1,0,0,1),\\
    P_C&=\mathrm{diag}(0,1,0,0,1,0),
\end{align}
For the effective Bloch eigenstate $|u_{n\mathbf k}\rangle$, the site weights are
\begin{equation}
    W_{\alpha,n}(\mathbf k)
    =
    \langle u_{n\mathbf k}|P_\alpha|u_{n\mathbf k}\rangle,
    \qquad
    \alpha=A,B,C .
\end{equation}
The normalized layer polarization plotted in Fig.~\ref{figS2} a is
\begin{equation}
    \mathcal P^L_n
    =
    \frac{
    W_{A,n}(\mathbf k)-W_{B,n}(\mathbf k)
    }{
    W_{A,n}(\mathbf k)+W_{B,n}(\mathbf k)
    },
    \label{eq:appD_layer_polarization}
\end{equation}
where states with negligible rim weight $W_A+W_B$ are treated as $C$-site dominated. Thus $\mathcal P^L_n=+1$ denotes an $A$-layer state and $\mathcal P^L_n=-1$ denotes a $B$-layer state, which are shown in Fig.~\ref{figS2} a. This normalized definition removes the trivial reduction of $W_A-W_B$ caused by admixture with the hub site.

The Berry curvature of band $n$ is evaluated from the Kubo formula
\begin{equation}
    \Omega_n(\mathbf k)
    =
    -2\,\mathrm{Im}
    \sum_{m\neq n}
    \frac{
    \langle u_{n\mathbf k}|\partial_{k_x}H_{\mathrm{eff}}|u_{m\mathbf k}\rangle
    \langle u_{m\mathbf k}|\partial_{k_y}H_{\mathrm{eff}}|u_{n\mathbf k}\rangle
    }{
    \left(E_{n\mathbf k}-E_{m\mathbf k}\right)^2
    },
    \label{eq:appD_band_berry}
\end{equation}
For an occupied subspace specified by the occupation factor $f_{n\mathbf k}$, the total Berry curvature is
\begin{equation}
    \Omega(\mathbf k)
    =
    \sum_n f_{n\mathbf k}\Omega_n(\mathbf k),
\end{equation}
The layer-resolved Berry curvature is obtained by weighting each band curvature by its layer weight,
\begin{align}
    \Omega_A(\mathbf k)
    &=
    \sum_n f_{n\mathbf k}
    W_{A,n}(\mathbf k)\Omega_n(\mathbf k),
    \nonumber\\
    \Omega_B(\mathbf k)
    &=
    \sum_n f_{n\mathbf k}
    W_{B,n}(\mathbf k)\Omega_n(\mathbf k),
    \nonumber\\
    \Omega_C(\mathbf k)
    &=
    \sum_n f_{n\mathbf k}
    W_{C,n}(\mathbf k)\Omega_n(\mathbf k).
    \label{eq:appD_layer_berry}
\end{align}
The distributions of Berry curvature with spin-valley-layer locking are shown in Fig.~\ref{figS2} b-d.

\bibliography{main}

\end{document}